# Monte Carlo computation of pair correlations in excited nuclei

Nicolas J. Cerf [1,2]

[1] *Division de Physique Théorique, Institut de Physique Nucléaire, Orsay Cedex 91406, France*
[2] *W. K. Kellogg Radiation Laboratory, 106-38, California Institute of Technology,
Pasadena, California 91125, USA*

(October 1995)

We present a novel quantum Monte Carlo method based on a path integral in Fock space, which allows to compute finite-temperature properties of a many-body nuclear system with a monopole pairing interaction in the canonical ensemble. It enables an exact calculation of the thermodynamic variables such as the internal energy, the entropy, or the specific heat, from the measured moments of the number of hops in a path of nuclear configurations. Monte Carlo calculations for a single-shell $(h_{11/2})^6$ model are consistent with an exact calculation from the many-body spectrum in the seniority model.

In the last decade, many advances in the solution of the nuclear many-body problem have become possible due to the development of quantum Monte Carlo (MC) methods. Standard Green's Function Monte Carlo (GFMC) methods have been used to perform a microscopic calculation of the ground state of nuclei up to $A = 6$ with realistic interactions [1,2]. Recently, a path-integral quantum Monte Carlo method has also been devised to explore the microscopic properties of nuclei at finite temperature [3]. This method, called Shell Model Monte Carlo (SMMC), uses the Hubbard-Stratonovich transformation of the many-body propagator to express observables as path integrals of one-body propagators in fluctuating auxiliary fields. It has been used successfully to treat shell-model spaces larger than those accessible to conventional methods [4–6]. In this Letter, we explore an alternative quantum MC method, based on a path integral in Fock space, that allows the calculation of thermodynamic properties of the many-body system. The observables are obtained through the MC computation of path integrals in Fock space, rather than in a space of auxiliary field variables as in SMMC, or in a spatial (and spin/isospin) coordinate space as in GFMC. One main advantage is that it yields a proper description of excited nuclei in terms of a canonical ensemble, without having to be concerned with particle-number projection. The method permits considering large model spaces, since it avoids an explicit enumeration of the many-body states. The results are in principle exact, i.e. they are subject only to controllable sampling errors. The time-discretization error, inherent to traditional path-integral MC methods (see e.g. [4,5]), is *absent* here due to the discrete nature of the states in Fock space.

An important motivation for an exact treatment of thermal properties of nuclei comes from the description of low-energy reactions, which depend crucially on level densities. In situations where the levels cannot be measured, as for example in the very neutron-rich nuclei involved in r-process nucleosynthesis, one still relies on semi-phenomenological level densities to calculate cross-sections [7]. In this Letter, we study a specific nuclear interaction, the pairing force, which exhibits the simplicity and efficiency of the proposed MC technique. It is well known that the finite-temperature BCS theory predicts a spurious phase transition due to particle-number fluctuations; this transition occurs at a temperature corresponding to an excitation energy of interest in many cases. Therefore, even if more refined methods have been devised such as the static-path approximation [8], an exact treatment is of a great help. The present work offers the perspective that, using an adequate procedure to circumvent the "sign problem" generic to fermionic systems [9], more general interactions could also be described by our path-integral MC method in Fock space; this is under investigation.

Consider a $N$-nucleon system with a monopole pairing interaction, described by the Hamiltonian

$$H = \sum_{k=1}^{\Omega} \epsilon_k (a_k^\dagger a_k + a_{-k}^\dagger a_{-k}) - \sum_{k,k'=1}^{\Omega} G_{k,k'} a_{k'}^\dagger a_{-k'}^\dagger a_{-k} a_k \quad (1)$$

where $k$ and $-k$ are time-reversed conjugate orbitals of energy $\epsilon_k$. Our MC simulation is based on a random walk in Fock space, where the two-body pairing interaction can be seen as a pair hopping term [10]; a pair of nucleons in orbits $(k, -k)$ hops to $(k', -k')$ with an amplitude $G_{k,k'}$. The set of basis states of our Fock space, i.e. configurations of $N$ nucleons into $2\Omega$ orbitals, are noted $C$ in the following. Because of the particular form of $H$, the hopping term does *not* act on unpaired nucleons; therefore, in this $C$-representation, one is considering a "mixed" system, where unpaired nucleons behave like "classical" particles, while pairs are "boson-like" particles (with a hard-core interaction preventing two pairs from occupying the same orbit pair). The thermodynamic properties of this system are determined from a path-integral MC computation of the density matrix. The decomposition of that matrix in $C$-space, $\rho(C, C') \equiv \langle C|\mathrm{e}^{-\beta H}|C'\rangle$, satisfies the so-called "sign rule", i.e. all matrix elements are non-negative, because the pairing interaction is attractive ($G_{k,k'} \geq 0$). As a consequence, this system can be efficiently simulated by quantum MC in $C$-space. In contrast, when treating more general interactions, paths of positive and negative sign that tend to cancel each other would contribute to the path integral in Fock space, thereby requiring to devise a specific solution to that "sign problem" for efficiently simulating the system.

In order to treat the operator $\mathrm{e}^{-\beta H}$ in $C$-representation, we work in interaction picture, writing $H = H_0 + H_1$ where $H_0$ and $H_1$ are the diagonal and off-diagonal part of $H$, respectively. We start with the Green's function $G(C, C'; t)$ in imaginary time ($\hbar = 1$), solution of the integral equation

$$G(t) = G_0(t) - \int_0^t \mathrm{d}t' G_0(t') H_1 G(t - t') \quad (2)$$

where $G_0 \equiv \mathrm{e}^{-tH_0}$, diagonal in $C$-space, plays the role of a "free-particle" propagator. The pair hopping term $H_1 = \sum_{k \neq k'} G_{k,k'} a_{k'}^\dagger a_{-k'}^\dagger a_{-k} a_k$ can be seen as an "interaction" Hamiltonian. Applying Eq. (2) recursively, as usual, yields an expansion of the full propagator in terms of number of pair hops

$$G(t) = G_0(t) - \int_0^t \mathrm{d}t_1 G_0(t_1) H_1 G_0(t - t_1) + \quad (3)$$

$$\int_{0 \leq t_1 \leq t_2 \leq t} \mathrm{d}t_1 \mathrm{d}t_2 G_0(t_1) H_1 G_0(t_2 - t_1) H_1 G_0(t - t_2) - \cdots$$

This expansion suggests one can perform integration over the paths in $C$-space with a MC simulation simply by sampling the times $t_1, t_2, \cdots$ at which pair hops occur, and the configurations $C$ in each inter-hop interval.

Consider now the MC calculation of thermodynamic averages in the canonical ensemble. The thermal expectation value of an observable $O$ at inverse temperature $\beta$, $\langle O \rangle_\beta \equiv \mathrm{Tr}(Oe^{-\beta H})/\mathrm{Tr}(e^{-\beta H})$, is derived by a stochastic calculation of the partition function $Z(\beta) = \mathrm{Tr}(e^{-\beta H})$. Using the expansion of the many-body Green's function $G(C, C'; t)$ in $C$-space, $Z(\beta)$ can be written as a path integral over chains of configurations, noted $C(t)$, periodic in time (of period $\beta$), having pair hops at times $t_1, t_2, \cdots t_D$. Here $D \equiv D[C(t)]$ is the number of hops [11] in the chain $C(t)$, and it can be any non-negative integer *excluding one*. It is the periodic boundary condition which forbids $D = 1$, and we will see below that it has important consequences. Note that, due to particle indistinguishability, periodic chains which make an overall pair permutation in one period have also to be considered. Thus, in order to calculate thermodynamic averages by MC, one has to sample the whole ensemble of chains of configurations $\{C(t)\}$ with variable $D \neq 1$ by use of the Metropolis algorithm [12]. That algorithm permits to sample a chain $C(t)$ with a probability proportional to its weight $W[C(t)]$ in $Z(\beta)$, given from Eq. (3) by

$$W[C(t)]dt_1 dt_2 \cdots dt_D = \exp\{A[C(t)]\} \prod_{i=1}^{D} G_{k,k'}(i) dt_i \quad (4)$$

Here $A[C(t)] = -\int_0^\beta E^{1b}(C(t))dt$ is the action in imaginary time corresponding to the one-body energy $E^{1b}$ of the configuration chain, and $G_{k,k'}(i)$ is the amplitude associated with the $i$-th hop along the chain. Since the propagator does not hop unpaired nucleons, the latter stay on the same orbit along the chain; their contribution to $E^{1b}$ is constant in time, simply giving an additive constant to the action.

Now, making use of the symmetry in imaginary time – all possible times $\tau$ at which the observable $O$ is inserted in the chain $C(t)$ when calculating $\langle O \rangle_\beta$ are equivalent – one can write thermal expectation values as averages over paths *and* over insertion times $\tau$. This gives $H[C(t)] = E[C(t)] - D/\beta$ as an estimator for the energy, with $E[C(t)] = \beta^{-1} \int_0^\beta E^{1b}(C(t))dt$ being the one-body energy averaged along the chain $C(t)$. An estimate of the internal energy is then obtained by

$$U \equiv \langle H \rangle_\beta - E_0 \simeq \langle\langle H[C(t)] \rangle\rangle_{C(t)} - E_0 \quad (5)$$

where $E_0$ stands for the ground state energy (extracted from the $\beta \to \infty$ limit) and $\langle\langle \cdot \rangle\rangle$ denotes a MC average over the sampled chains $C(t)$. Thus, $U$ is a function of the average number of hops along the chain $\langle\langle D \rangle\rangle$ (any additional hop gives on average the contribution $-\beta^{-1}$ to $U$). The same reasoning yields an estimate of the pair occupation probability in the $k$-th orbital, $\langle n_k \rangle_\beta \simeq \langle\langle n_k[C(t)] \rangle\rangle_{C(t)}$, with $n_k[C(t)] = \beta^{-1} \int_0^\beta n_k(C(t))dt$. Thus, the occupation of the $k$-th orbital, averaged along each chain $C(t)$ and averaged over the sampled chains, simply yields the exact quantum expectation value of $n_k$ at inverse temperature $\beta$. One can also obtain an estimator of the specific heat from the variance of the energy, $C = \beta^2 \mathrm{Var}\{H\}$, which is given by

$$C \simeq \beta^2 \mathrm{Var}\{H[C(t)]\}_{C(t)} - \langle\langle D \rangle\rangle_{C(t)} \quad (6)$$

Thus, the specific heat depends on the variance of the number of hops along the chain, and on the correlation between the number of hops $D$ and the averaged one-body energy of a chain $E[C(t)]$. Finally, the entropy $S$ in the canonical ensemble can be estimated by use of MC (see below). Then, the level density can be calculated via the saddle-point approximation as $\rho(U) = \beta e^S/(2\pi C)^{1/2}$, using the MC estimates for $S$ and $C$. This provides a natural extension to the MC approach to level densities for non-interacting nucleons suggested in [13].

A general difficulty related to the Metropolis method is to find an efficient algorithm for generating trial moves (*i.e.*, for creating a new $C(t)$ from the current one), which has a non-negligible acceptance rate and a reasonably short auto-correlation time. The detail of a general algorithm for sampling the paths $C(t)$ will be reported elsewhere. For illustrative purposes, we focus in this Letter to the case of a single-shell model ($\epsilon_k = 0$) with a constant pairing strength $G$. This model, called the seniority model [14], can been solved exactly using quasi-spin formalism, allowing us to demonstrate the interest of the MC procedure. Defining the seniority quantum number $s$ ($s = 0, 2, \cdots N$ for even $N$, only considered here), the $N$-body energy levels are $E(N, s) = -G(N - s)(2\Omega - s - N + 2)/4$ with a degeneracy $d(s) = \binom{\Omega}{s/2} - \binom{\Omega}{s/2-1}$; note $d(0) = 1$. This model is simpler to compute by MC than a general case because $E^{1b}(C) = -NG/2$ is independent of $C$, so that the weight of any chain having $D$ hops scales as $G^D$. Also, integration over (time-ordered) intermediate times $t_1, t_2 \cdots t_D$ can be made explicitly (without requiring MC calculation), giving the factor $\beta^D/D!$ in the weigths, so that it is enough to sample periodic sequences of $C$'s of variable length $D$ ($D = 0, 2, 3, \cdots$). In those sequences $\{C_1, C_2, \cdots C_D\}$, $C_i$ and $C_{i+1}$ must differ by exactly one hop.

As an illustration, we consider a half-filled $h_{11/2}$ shell with a constant pairing strength $G = 1$ MeV. We plot in Fig. 1 the thermal expectation value of energy as a function of inverse temperature. The energy in the seniority model is estimated by MC from $-NG/2 - \beta^{-1} \langle\langle D \rangle\rangle$. The squares are the result of MC simulations using a sample of $10^6$ sequences, while the solid line was calculated numerically from the exact spectrum in the seniority model.

FIG. 1. Temperature dependence of the expectation value of the many-body energy, $\langle H \rangle_\beta$, in a single-shell $(h_{11/2})^6$ model with a constant pairing strength $G = 1$ MeV. The squares represent Monte Carlo results at each inverse temperature $\beta = T^{-1}$. The solid curve corresponds to an exact numerical calculation from the spectrum in the seniority model. In the inset, we show the average number of hops $\langle\langle D \rangle\rangle$ derived from the MC simulations (squares) as a function of the chain length $\beta$. The dashed curve, corresponding to the asymptotic $(\beta \to \infty)$ behaviour $\langle\langle D \rangle\rangle = \mu = \beta G N(2\Omega - N)/4$, illustrates the hindrance due to periodic boundary conditions.

FIG. 2. Temperature dependence of the entropy, $S$, calculated from $S = \ln Z(\beta) + \beta U$ in the same model. The Monte Carlo calculations correspond to squares, while the exact calculation is shown as a solid curve. In the inset, we show the corresponding results for the specific heat $C$, calculated by a finite difference approximation to $\partial U / \partial T$. The observed peak is a clear signature of the vanishing of pair correlations at a temperature around $T \simeq 2$ MeV.

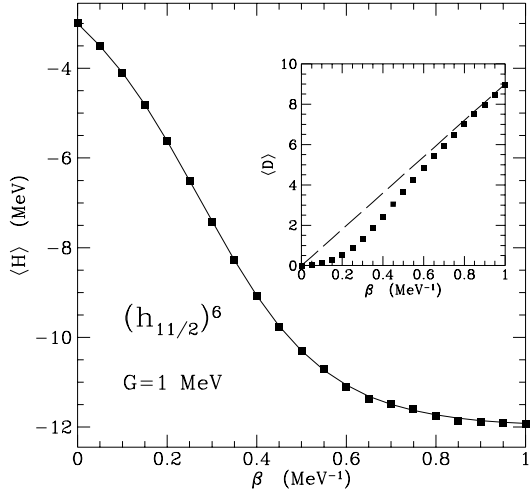

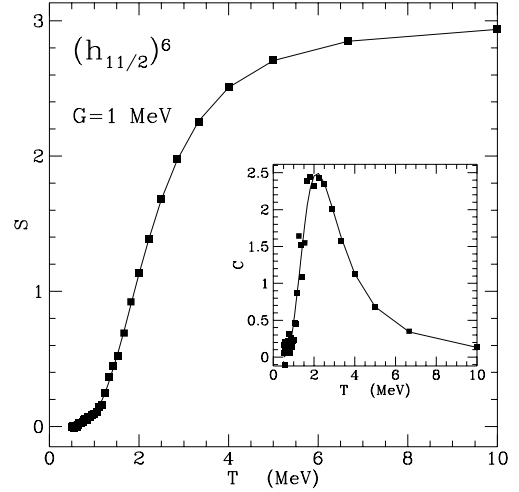

The agreement is very good. The inset in Fig. 1 shows the corresponding average number of hops $\langle\langle D \rangle\rangle$ in the chains sampled by MC (squares) as a function of the chain length $\beta$. For long chains, $\langle\langle D \rangle\rangle$ increases linearly with the chain length, with the asymptotic $(\beta \to \infty)$ behaviour $\langle\langle D \rangle\rangle = \beta G N(2\Omega - N)/4$ (dashed line). This is expected since it was shown in ref. [10] that, at zero temperature, the number of hops [15] per imaginary-time unit $\Delta t$ is drawn from a Poisson distribution of parameter $\nu = \Delta t G N(2\Omega - N)/4$. For short chains, however, $\langle\langle D \rangle\rangle$ increases quadratically, as $\beta^2 G^2 N(2\Omega - N)/4$, and there is a depletion in the average number of hops. That depletion, which reflects the hindrance due to periodic boundary conditions ($D = 1$ is excluded), implies the non-zero specific heat in this model as shown below.

In Fig. 2, we plot the entropy $S$ as a function of temperature $T = \beta^{-1}$. The Monte Carlo calculation of $S$ in the canonical ensemble (squares) uses relation $S = \ln Z(\beta) + \beta U$, where $\ln Z(\beta)$ is obtained by numerically integrating $U$ from 0 to $\beta$ (remember $\partial \ln Z/\partial \beta = -U$), using $S(0) = \ln \binom{\Omega}{N/2}$. The agreement with the exact numerical calculation of $S$ (solid line) derived from the spectrum is evident. The inset in Fig. 2 shows the temperature dependence of the specific heat $C$ in this model. Since $E[C(t)] = -NG/2$ in the seniority model, the MC estimator is simply $C \simeq \text{Var}\{D\} - \langle\langle D \rangle\rangle$. According to ref. [10], for $\beta \to \infty$, i.e. when the effect of periodic boundary conditions vanishes, the number of hops $D$ in the chain of length $\beta = M\Delta t$ is distributed as a Poisson variable of parameter $\mu = M\nu = \beta G N(2\Omega - N)/4$; then, it is obvious that $C$ tends to zero for $\beta \to \infty$, as expected. An independent method to access to the specific heat by MC is to make use of $C = \partial U/\partial T = -\beta^2 \partial U/\partial \beta$. Graphically, it implies that the tangent of the curve $\langle\langle D \rangle\rangle$ versus $\beta$ (see inset of Fig. 1) intercepts the y-axis at $-C$; therefore, the depletion of that curve compared to the asymptotic curve indicates that $C$ tends to a maximum value and then falls to zero. The squares, obtained by MC (by numerically estimating the derivative of $U$), are in good agreement with the solid line, calculated from the exact spectrum. The MC determination of $C$ from the energy variance (not shown in Fig. 2) is also compatible with the exact result, but is affected by a much larger statistical noise for large $\beta$. The peak in the specific heat around $T \simeq 2$ MeV reflects the pairing phase transition; above that temperature, the pair correlations vanish. The MC simulations also exhibit a collapse in the measured pairing field $\langle \Delta^\dagger \Delta \rangle$ near 2 MeV, with $\Delta = \sum_k a_{-k} a_k$, which is a clear signature of the pairing transition.

In conclusion, we have shown that a path-integral Monte Carlo method in Fock space is well-suited to describing pair correlations in nuclei at finite temperature. It allows a stochastic calculation of thermodynamic properties (internal energy, entropy, specific heat, level density) in the canonical ensemble; there is no need to project into the number of particles. The scattering of pairs in the orbits is simulated straigthforwardly; for instance, the pair occupation probabilities are measured during the simulation, *not* calculated. Estimators for thermal expectation values of observables have simple expressions depending on the number of pair hops $D$ in the path; for instance, the internal energy (or specific heat) is related to the average (or variance) of $D$. Results for a single-shell $(h_{11/2})^6$ model with a constant pairing strength of 1 MeV are compared with a numerical calculation from the exact spectrum in order to ascertain the method. The observed peak in the specific heat is associated with the vanishing of nucleon pair correlations at a temperature of $\simeq 2$ MeV. In contrast to conventional path-integral MC calculations which need to discretize the imaginary time into time slices $\Delta t$ (see *e.g.* [4,5]), our MC method uses a continuous-time formalism; this is possible because the path integral lies in a discrete configuration space. Therefore, there is no time-discretization error (or no need to extrapolate to $\Delta t \to 0$) and the *only* residual error in our MC method is the statistical noise, which can be made arbitrarily small. The extension of these MC calculations to a more general pairing interaction, or even to other interactions, should allow an exact description of nuclear properties at high excitation energies.

## ACKNOWLEDGMENTS

We are grateful to S. E. Koonin for initially motivating this work, and to O. C. Martin for many helpful discussions. The author is supported by a grant from the Human Capital and Mobility Program of the European Union. The Division de Physique Théorique is Unité de Recherche des Universités Paris XI et Paris VI associée au CNRS.